# Application of Deep Q-Network in Portfolio Management

Ziming Gao, Yuan Gao, Yi Hu, Zhengyong Jiang†, Jionglong Su†,
Department of Mathematics, Xi'an Jiaotong-Liverpool University
Suzhou, P.R.China
†Zhengyong.Jiang@xjtlu.edu.cn   †Jionglong.Su@xjtlu.edu.cn

*Abstract*—Machine Learning algorithms and Neural Networks are widely applied to many different areas such as stock market prediction, face recognition and population analysis. This paper will introduce a strategy based on the classic Deep Reinforcement Learning algorithm, Deep Q-Network, for portfolio management in stock market. It is a type of deep neural network which is optimized by Q Learning. To make the DQN adapt to financial market, we first discretize the action space which is defined as the weight of portfolio in different assets so that portfolio management becomes a problem that Deep Q-Network can solve. Next, we combine the Convolutional Neural Network and dueling Q-net to enhance the recognition ability of the algorithm. Experimentally, we chose five low-relevant American stocks to test the model. The result demonstrates that the DQN based strategy outperforms the ten other traditional strategies. The profit of DQN algorithm is 30% more than the profit of other strategies. Moreover, the Sharpe ratio associated with Max Drawdown demonstrates that the risk of policy made with DQN is the lowest.

## I. INTRODUCTION

Optimized stock trading strategy is a process of making decisions based on optimizing allocation of capital into different stocks in order to maximize performance, such as expected return and Sharpe ratio. Traditionally, there exist portfolio trading strategies which may be broadly classified into four categories, namely "Follow-the-Winner", "Follow-the-Loser", "Pattern-Matching", and "Meta-Learning" [1], [7]. However, in financial environments, there exist correlations between the price and other factors, as well as substantial noise, rendering the traditional methods to be limited in their use. In view of this, deep machine-learning approaches are now applied to financial market trading [23]. Nevertheless, many of them tend to predict price movements by inputting history asset prices to output a prediction of asset prices in the next trading period via neural network, and the trading agent will take action based on these predictions [1], [8], [9]. This idea seems reasonable but the performance of these algorithms is highly dependent on the prediction accuracy of future market prices. Therefore, many studies [24],[25] solve the problem by Reinforcement Learning without predicting future prices. Previous studies indicated good performance in each setting, though, there still exist limitations such as it is not adaptable to multi-asset portfolio [1]. Learning good policies that bring more profit by optimizing an accumulative future reward signal in sequential decision-making problems is the goal of Reinforcement Learning, with Q Learning being one of the most popular Reinforcement Learning algorithms [2]. More recently, Deep Q-Networks algorithm which combines Q Learning with deep neural network is introduced and applied to many areas. In this paper, we propose a DQN framework specifically designed for a multi-asset portfolio trading strategy. This framework allows the DQN agent to optimize trading strategies through learning from its experience in financial environment so that it can adapt to real financial market. In our study, a discrete action space is defined to increase practicality of strategies, and algorithm performance is also improved by taking advantage of double DQN [2] and dueling DQN [5].

The rest of this paper is organized as follows. Section II defines the portfolio management problem that this project is aiming to solve, and all the assumptions of this study are listed in section III. Section IV introduces discrete actions, and asset preselection and input price tensor are shown in Section V. Section VI and Section VII presents DQN algorithm, and Section VIII shows network topology. Experiment and result are stated in Section IX. Section X contains conclusions and future work.

## II. PROBLEM STATEMENT

In classical portfolio management theory, portfolio management aims to find the best investment policy that gives the maximum overall portfolio in a given period. In practice, the investor modifies the weight of portfolio in different assets according to the price of these assets and the previous distribution of portfolio, and this process is approximately described as Markov Decision Process (MDP) [6]. Essentially, MDP is a mathematical model that is used to formulate optimized policy, which consists of a tuple $(S_t, a_t, P_t, R_t)$. The meaning of each element in the tuple is listed below

- $S_t$ - the state at time $t$
- $a_t$ - the action taken at time $t$
- $P_t$ - the probability of transforming the state from $S_t$ to $S_{t+1}$
- $R_t$ - reward at time $t$

We can construct a model for portfolio management problem by defining the state at time $t$, $S_t$, to be the price of assets invested, and action of time $t$ as

$$a_t \triangleq w_{t+1} - w_t \quad (1)$$

where $w_t$ and $w_{t+1}$ are the weight vectors of portfolio at time $t$ and $t+1$ respectively. Furthermore, we define the reward as

$$R_t \triangleq p_{t+1} - p_t \quad (2)$$

where $p_t$ is the portfolio at time $t$ and $p_{t+1}$ is the portfolio at time $t+1$.

In (2), we only think of the reward at the present time $t$, but with a given policy $\pi$, the state at time $t$ affects the all the states after time $t$, which means the value of $S_t$ is not only $R_t$, but also the rewards of following time periods. Therefore, with policy $\pi$, the value function $G_\pi$ of $S_t$ should be defined as

$$G_\pi(S_t) \triangleq \sum_{k=t}^{T} \gamma^{k-t} R_k \quad (3)$$

in which $T$ is the last trading period and $\gamma \in (0, 1]$ is a discount factor. In general, $G_\pi(S_t)$ cannot simply be obtained using (3), so we need to compute its estimated value by taking the expectation of $G_\pi$. And since the policy $\pi$ is determined uniquely by action $a_t$, we define the value function $Q_\pi$ of $S_t$ and $a_t$ as following

$$Q_\pi(S_t, a_t) \triangleq E[G_\pi(S_t)] \quad (4)$$

Based on (4), we may estimate the value of a state and an action that can be taken on this state, which is the basic principle of Q Learning.

Deep Q-Network is an improvement over classic Q learning [3]. In Q Learning, we obtain the state $S_t$ and the action $a_t$ of time $t$ as its input and compute Q-value, $Q_\pi(S_t, a_t)$, as its output. By searching all the possible combinations of states and actions, we can obtain a Q-table of which the rows are states and columns are actions. From this table, we know the best action for each state so that the optimized policy may be determined. However, considering that Q Learning requires all the possible combinations of states and actions to be explicitly known, it is not suitable to solve problems with infinite state space, such as portfolio management in stock market. Therefore, we use Deep Q-network as our basic model which does not have special requirement for state space, i.e., it can solve either finite or infinite state space. According to [3], DQN receives a state $S_t$ as the input and output Q-value $Q(S_t, a)$ of each action $a \in A$, where $A$ is a finite set.

Based on the Deep Q-network theory, this reinforcement learning algorithm uses neural networks to obtain the value of each action, which can be applied to solve portfolio management problems that have infinite state spaces. Nevertheless, it still requires a finite action space. Considering the actions which can be taken in stock market are infinite, we shall define a new discrete action space so that DQN may be applied to this financial market.

III. ASSUMPTIONS

Before introducing the model, we shall make several assumptions to simplify our problem
- Assumption 1: The action taken by the agent will not affect the financial market
- Assumption 2: The portfolio remains unchanged between the end of previous trading period and the beginning of next trading period
- Assumption 3: There is no other assets that can be chosen besides the selected assets
- Assumption 4: Since the proportion of commission fee is only 0.0025, we approximately set it zero, which means the portfolio value will not be reduced when it is reallocated.
- Assumption 5: The volume of each stock is large enough, so the agent can buy or sell each of them at any trading day.

IV. ACTION DISCRETIZATION

Traditionally, the action in portfolio management is defined as (1), which is the difference between $w_{t+1}$ and $w_t$. However, this may lead to very complicated action space or even continuous action space. Therefore, we redefine action as exactly the weight

$$a_t \triangleq w_t \quad (5)$$

Based on the definition, we create an environment for DQN algorithm so that it can adapt to stock market, which can be started with discretizing the action space.

To begin with, we regard the initial portfolio as 1, and equally divide it into $N$ parts. By doing this, we obtain the smallest unit of the portfolio, which is $1/N$. Then, for total $M + 1$ assets (including cash), we can calculate each action by using permutation. This process is equivalent to placing $N$ balls into $M+1$ baskets. In (6), the different combinations that make up the action space are shown. Refer to Algorithm. 1 to understand how this may be achieved.

$$\begin{pmatrix} \frac{N}{N}, 0, 0, 0, \dots, 0 \end{pmatrix}$$
$$\begin{pmatrix} \frac{N-1}{N}, \frac{1}{N}, 0, 0, \dots, 0 \end{pmatrix}$$
$$\begin{pmatrix} \frac{N-1}{N}, 0, \frac{1}{N}, 0, \dots, 0 \end{pmatrix}$$
$$\vdots$$
$$\begin{pmatrix} \frac{N-1}{N}, 0, 0, 0, \dots, \frac{1}{N} \end{pmatrix}$$
$$\vdots$$
$$\begin{pmatrix} 0, 0, 0, 0, \dots, \frac{N}{N} \end{pmatrix}$$
(6)

```
Algorithm 1 Action Discretization Algorithm
Input: Asset number M + 1, number of division N
Output: Action space A
1:  itemNUM = (M + 1) + N - 1
2:  pointer = 0
3:  seq = (0, 1, 2, ..., itemNUM - 1)
4:  for c ∈ combinations(seq, M) do
5:      action = (0, 0, ..., 0)
6:      for i = 0 → len(c) - 1 do
7:          action(i + 1) = c(i + 1) - c(i) - 1
8:      end for
9:      action(0) = c(0)
10:     action(M) = itemNUM - c(M - 1) - 1
11:     for j = 0 → M do
12:         action(j) = action(j)/N
13:     end for
14:     A(pointer) = action
15:     pointer = pointer + 1
16: end for
```

Algorithm 1. Action Discretization Algorithm

In Algorithm. 1, *combinations(seq, M)* refers to the set of combinations consisting of $M$ elements from *seq*. So far, based on the least unit $(1/N)$ and the combination algorithm mentioned above, we can discretize the action space, and the total number of actions is $\binom{M+N-1}{M-1}$. Here, since $M$ and $N$ are both positive integers, the action space is finite.

## V. MATHEMATICAL FORMALISM

### A. Data Processing

The data processing method is inspired by [1]. The state received by our Deep Q-Network consists of the portfolio weight vector at the beginning of the last trade period, $w_{t-1}$, and the price tensor $P_t$ that includes closing, opening, highest and lowest price for the assets in the previous $n$ days. That is, the state at trade period $t$, is defined as the following 2-tuple

$$S_t \triangleq (P_t, w_{t-1})$$

$$s.t. \begin{cases} w_{t-1} = [w_{t-1,0}, w_{t-1,1}, ..., w_{t-1,M}] \\ P_t = [P_t^o, P_t^c, P_t^h, P_t^l] \end{cases} \quad (7)$$

where $w_{t-1,i}$ denote the proportion of $i$-th asset (Here, 0-th asset is cash) at the beginning of the last trade period and $\sum_{i=0}^{M} w_{t-1,i} = 1$. Meanwhile, according to [1], we set the initial weight as

$$w_0 = [1, 0, 0, ..., 0] \quad (8)$$

where $w_0$ is a $1 \times (M + 1)$ vector. For the price tensor $P_t$, it is transformed from the original price tensor $P_t^*$ consisting of $P_t^o, P_t^c, P_t^h, P_t^l$ which are the normalized price matrices of opening, closing, highest and lowest price which are denoted as below

$$P_t^o = [p_{t-n+1}^o \oslash p_t^c | ... | p_t^o \oslash p_t^c] \quad (9)$$

$$P_t^c = [p_{t-n+1}^c \oslash p_t^c | ... | p_t^c \oslash p_t^c] \quad (10)$$

$$P_t^h = [p_{t-n+1}^h \oslash p_t^c | ... | p_t^h \oslash p_t^c] \quad (11)$$

$$P_t^l = [p_{t-n+1}^l \oslash p_t^c | ... | p_t^l \oslash p_t^c] \quad (12)$$

where $\oslash$ is elementwise division. In addition, $p_t^o, p_t^c, p_t^h, p_t^l$ represent the price vectors of opening, closing, highest and lowest price for all assets in trade period $t$ respectively. In other words, the $i$-th element of them, $p_{t,i}^o, p_{t,i}^c, p_{t,i}^h, p_{t,i}^l$, are relative technical indicators of $i$-th asset in the $t$-th period. Therefore, if there are $M$ assets (except cash) in the portfolio, the original price tensor $P_t^*$ is an *(M, N, 4)*-dimensional tensor, as Fig. 1

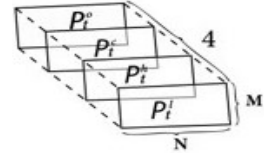

Figure 1. Original Price tensor $P_t^*$

We note that simply normalizing the original data may cause some recognition problem, i.e., the data may not provide enough information for the network to distinguish. Considering that, we need to subtract each element in $P_t^*$ by 1 and multiply it by an *expansion coefficient* $\alpha$. So the final price tensor is defined as

$$P_t \triangleq \alpha(P_t^* - 1) \quad (13)$$

where **1** is a tensor in dimension of *(M, N, 4)*, and with all elements as 1.

### B. Interaction with Environment

With the previous definition of state and action, we can define the transitions in the financial market environment. Since the opening price in period $t$ is not equal to the closing price in period $t$, we denote the original price relative vector of $t$-th trading period, $\boldsymbol{\mu}_t^*$, to be

$$\boldsymbol{\mu}_t^* = \boldsymbol{p}_t^c \oslash \boldsymbol{p}_t^o = \left(\frac{p_{t,1}^c}{p_{t,1}^o}, \frac{p_{t,2}^c}{p_{t,2}^o}, \dots, \frac{p_{t,m}^c}{p_{t,m}^o}\right) \quad (14)$$

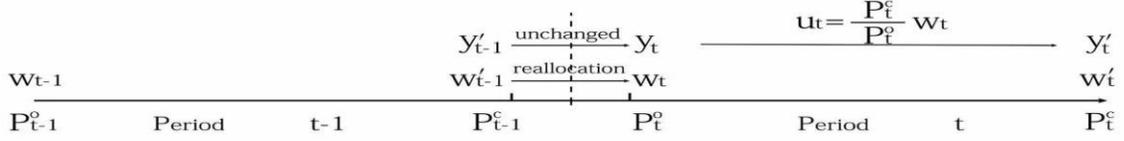

Figure 2. Trading Process

Here, since the total portfolio value includes cash, so we need to add cash price to $\boldsymbol{\mu}_t^*$. Considering cash price remains unchanged, $\boldsymbol{\mu}_t$ would take the following form

$$\boldsymbol{\mu}_t = \left(1, \frac{p_{t,1}^c}{p_{t,1}^o}, \frac{p_{t,2}^c}{p_{t,2}^o}, \dots, \frac{p_{t,m}^c}{p_{t,m}^o}\right) \quad (15)$$

Therefore $y_t'$, the portfolio value at the end of period $t$, can be computed by

$$y_t' = y_t \boldsymbol{w}_t \cdot \boldsymbol{\mu}_t \quad (16)$$

where $y_t$ denote the portfolio value at the beginning of $t$-th trade period and $\boldsymbol{w}_t$ is asset weight vector at the beginning of current trading period. Furthermore, since commission fee is zero (Section III, Assumption 4), $y_{t-1}'$ equals $y_t$, and thus we define the rate of return $\rho_t$ as

$$\rho_t \triangleq \frac{y_t'}{y_{t-1}'} - 1 = \frac{y_t'}{y_t} - 1 = \boldsymbol{w}_t \cdot \boldsymbol{\mu}_t - 1 \quad (17)$$

Defining the rate of return in this way is reasonable, but our primary goal is to maximize the overall portfolio. So the reward $r_t$ is defined as

$$r_t \triangleq \ln\left(\frac{y_t'}{y_{t-1}'}\right) = \ln(\boldsymbol{w}_t \cdot \boldsymbol{\mu}_t) \quad (18)$$

We can thus compute the total portfolio at the end of the trading time by summing $r_t, t = 1,2,\dots,T$ together and taking exponential

$$\exp(\sum_{t=1}^T r_t) = \exp\left(\sum_{t=1}^T \ln\left(\frac{y_t'}{y_{t-1}'}\right)\right) = \frac{y_T'}{y_0'} \quad (19)$$

The result obtained in (19) is a fraction, so if we want to compute the final portfolio $y_T'$, we need to multiply the result by the initial portfolio $y_0'$

$$y_T' = \frac{y_T'}{y_0'} \cdot y_0' = \exp\left(\sum_{t=1}^T r_t\right) \cdot y_0' \quad (20)$$

In summary, the transition of stock environment is demonstrated by Fig. 2 and the goal of our algorithm is to maximize the final portfolio (20) during a given trading time.

## VI. PRIORITIZED SAMPLING

Traditional DQN takes samples from the its memory pool randomly, which is not efficient. More importantly, if the good samples are very difficult to obtain, which means the samples that are valuable to be learnt are very rare in the memory pool, the agent will hardly learn anything useful. However, in financial market, it is very hard to obtain such valuable memory due to the complexity of market variation. Therefore, we need to use more efficient sampling method, which is called Prioritized Experience Replay [4], to enhance the learning ability of the agent.

According to [4], Prioritized Experience Replay takes samples from memory pool by *TD-error*, which is defined as

$$\text{TD-error} \triangleq |Q_{real} - Q_{eval}| \quad (21)$$

here, $Q_{eval}$ is the output of evaluation network (section VII), and $Q_{real}$ is defined in (29).

In other words, with this sampling method, the memory with larger *TD-error* are more likely to be chosen, which means the agent will learn those experience of which the difference between predicted value and real value is large first. By doing this, the agent will learn more valuable experience from the training process, and also enhance its adaptation for some extreme cases, such as sudden increase or decrease of price.

However, considering the huge memory size and difficulty in searching the proper sample, we need to introduce a structure called *SumTree* [4], and its basic structure is shown in Fig. 3. Notice here, this tree structure only stores the *TD-errors* of each memory and their sum. The memories are stored in memory pool which is another container.

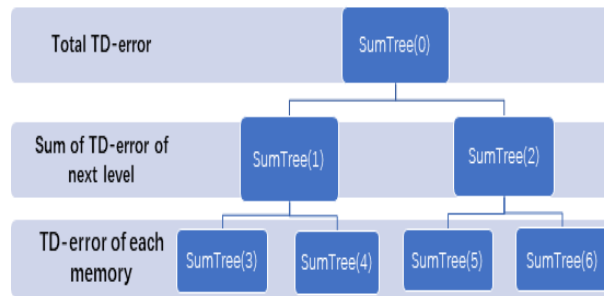

Figure 3. SumTree

From this structure, we can easily compute the total *TD-error* of all the samples in memory pool by adding all the *TD-error* in the bottom nodes together, and its tree-shape is very helpful for searching. The searching algorithm is shown in Algorithm. 2.

```
Algorithm 2 SumTree Searching Algorithm
Input: Total TD-error p, total node of SumTree M
Input: Memory size Z, batch size N
Output: Sample TD-errors T, samples S
 1: seg = p/N
 2: for i = 0 → N − 1 do
 3:     a, b = seg * (i), seg * (i + 1)
 4:     v = uniform(a, b)
 5:     parentIDX = 0
 6:     search = True
 7:     while search is True do
 8:         leftIDX = 2 * parentIDX + 1
 9:         rightIDX = leftIDX + 1
10:         if leftIDX >= M then
11:             leafIDX = parentIDX
12:             search = False
13:         else
14:             if v < SumTree(leftIDX) then
15:                 parentIDX = leftIDX
16:             else
17:                 v = v − SumTree(leftIDX)
18:                 parentIDX = rightIDX
19:             end if
20:         end if
21:     end while
22:     dataIDX = leafIDX − Z + 1
23:     T(i) = SumTree(leafIDX)
24:     S(i) = Memory(dataIDX)
25: end for
```

Algorithm 2. SumTree Searching Algorithm

In Algorithm 2, *uniform(a, b)* refers to sampling from a uniform distribution defined on *(a, b)*, and *SumTree* is the structure mentioned in Fig. 5, *Memory* is the memory pool that stores memories *($S_t$, $a_t$, $r_t$, $S_{t+1}$)* and the capacity of memory pool equals the number of nodes at the bottom level of the *SumTree*.

Additionally, in the searching process, we also need to calculate the weight for each sample, which will be used to calculate the loss function (32). The weight for sample *i* is given by the following equation

$$weight_{(i)} = \left(p_{(i)}/p_{min}\right)^{-\beta}; i = 1,2,\dots,n \quad (22)$$

Where $p_{(i)}$, $p_{min}$ are the *TD-error* of sample *i* and minimum *TD-error* for all experiences in the memory pool respectively, and $\beta \in (0,1]$ is a constant. Therefore, by Prioritized Experience Replay, we can obtain a set of valuable samples and a weight vector **K** which is defined as

$$\boldsymbol{K} = [weight_{(1)}, \dots\dots, weight_{(n)}] \quad (23)$$

where *n* is the batch size.

## VII. TRAINING PROCESS

Considering that the basic principle of DQN is to approximate the real Q-function, there should be two Deep Q-Networks, i.e., the evaluation network $Q_{eval}$ and the target network $Q_{target}$, which have exactly same structure, but different parameters. The parameters of $Q_{eval}$ are continuously updated, while the parameters of $Q_{target}$ are fixed until they are replaced by the parameters of $Q_{eval}$.

In the training process, the sampler will take a batch of memories

$$\{(S_{t_1}, a_{t_1}, r_{t_1}, S_{t_1+1}), \dots\dots, (S_{t_n}, a_{t_n}, r_{t_n}, S_{t_n+1})\} \quad (24)$$

from the memory pool according to Prioritized Experience Replay (Section VI). The $Q_{eval}$ receives the present state $S_t$ as input and returns the estimated Q-values $Q_{eval}(S_t, a)$ for each action $a \in A$. Simultaneously, $Q_{target}$ receives the state of next time $S_{t+1}$ and also returns the Q-values $Q_{target}(S_{t+1}, a)$ for each action $a \in A$. Then, according to the theory of double DQN, we select Q-values from $Q_{target}$ and $Q_{eval}$ by the following equations

$$Q_{target^*(i)} = Q_{target}\left(S_{t_i+1}, argmax\left(Q_{eval}(S_{t_i+1}, a)\right)\right) \quad (25)$$

$$Q_{eval^*(i)} = Q_{eval}\left(S_{t_i}, argmax\left(Q_{eval}(S_{t_i}, a)\right)\right) \quad (26)$$

Where $i = 1, 2, \dots, n$ and we obtain the vector of target Q-values $\boldsymbol{Q}_{target^*}$ and the vector of estimated Q-values $\boldsymbol{Q}_{eval^*}$ by

$$\boldsymbol{Q}_{target^*} = [Q_{target^*(1)}, Q_{target^*(2)}, \dots, Q_{target^*(n)}] \quad (27)$$

$$\boldsymbol{Q}_{eval^*} = [Q_{eval^*(1)}, Q_{eval^*(2)}, \dots, Q_{eval^*(n)}] \quad (28)$$

Then obtain the real Q-value and the vector of real Q-value by

$$Q_{real(i)} = r_{t_i} + \gamma Q_{target^*(i)}; i = 1,2,\dots,n \quad (29)$$

$$\boldsymbol{Q}_{real} = [Q_{real(1)}, Q_{real(2)}, \dots, Q_{real(n)}] \quad (30)$$

where $\gamma \in (0, 1]$. And we define the original loss function as

$$l^* = (\boldsymbol{Q}_{eval^*} - \boldsymbol{Q}_{real}) \odot (\boldsymbol{Q}_{eval^*} - \boldsymbol{Q}_{real}) \quad (31)$$

Here, $\odot$ is elementwise product. Since we use Prioritized Experience Replay (Section VI), so each sample in the minibatch has different weight which has been defined in (22). Therefore, the final loss should be in the following form

$$l = \boldsymbol{K} \cdot l^* \quad (32)$$

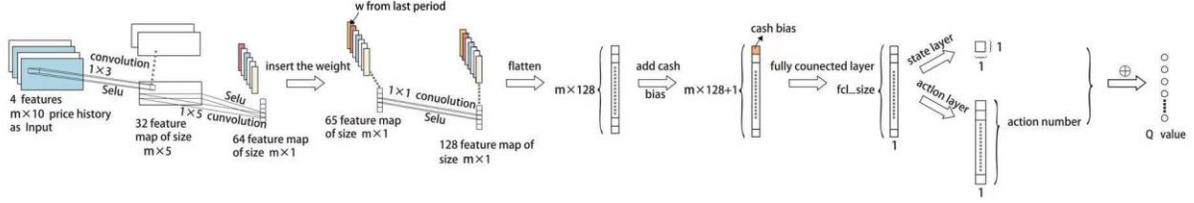

Figure 4. Network Topology

where $K$ is the weight vector defined in (23). Once the loss function is defined, we can train the network by minimizing the loss function.

## VIII. NETWORK TOPOLOGY

This network is based on Convolution Neural Network (CNN), and the specific topological structure is inspired by [1]. Firstly, we set the input of this network as 2-tuple

$$I = (P_t, w_{t-1}) \qquad (33)$$

which means the agent receives price data of trade period $t$, $P_t$ (defined in (13)) and portfolio weight of trade period $t-1$. Notice here, although the weight of previous trading period is put into the network, it will not go through the first two convolution layers. Moreover, we choose *Selu* as activation function of the convolutional layers, because there are a number of negative numbers in the processed data defined in (13), and *Relu* will transform all the negative numbers to zero, which causes serious death of neurons.

After the network receives the input, the price tensor $P_t$ will go through the first convolution layer of which the kernel is $1 \times 3$, and this layer will output 32 feature maps with size of $m \times 5$. In the second convolution layer, the kernel is $1 \times 5$ and output 64 features with size $m \times 1$. Then, insert the weight of previous trading period $w_{t-1}$ in the feature map and take this 65 features as the input of the next convolution layer (Here, the first item of $w_{t-1}$, which is the weight of cash, is removed). In the third convolution layer, the kernel is $1 \times 1$ and 128 features are taken. After the third convolution layer, the 128 feature maps will be flattened and a cash bias will be added. So far, we obtain the state of trading period $t$, which can be expressed as

$$S_t = c_3 \circ c_2 \circ c_1(P_t, w_{t-1}) \qquad (34)$$

In (34), we regard each convolutional layer as a function and $c_i$ represents $i$-th convolutional layer, with ∘ as composition of functions.

Next, the part of the network after convolutional layers can be regarded as a Q-net, which receives the state $S_t$ and output Q-values. What should be explained detailly is the structure of the second fully connected layer. Here, we use the structure called dueling Q-net, which was introduced by [5]. By this structure, the network can evaluate the value of state and the value of action separately, which helps the agent to evaluate the current situation completely and also take an action wisely. Finally, the Q-value of $S_t$ is given by

$$Q_{(S_t,a)} = Q_s + (Q_a - E[Q_a]) \qquad (35)$$

where $Q_s$ is the output of state layer, and $Q_a$ is the output of action layer.

## IX. EXPERIMENT

### A. Experimental Setting

In our experiment, five low-relevant US stocks from Yahoo Finance[1] had been chosen as risk assets and code of them are CAH, CAT, CCE, CCL, DIS. Together with the cash as risk-free asset, there were 6 investment products to be managed. We set the trade period as two days in order to increase the difference between tensors. Meanwhile, we selected the past 3 years as a training set and back testing period, with 2015/01/02 - 2016/12/30 as the period of training set and 2017/01/05 - 2017/11/17 as the period of back testing set.

### B. Performance Metrics

Three different metrics had been used to evaluate the performance of trading strategies. The first metric is accumulative rate of return [1], defined as

$$ARR = \exp(\sum_{t=1}^{T} r_t) - 1 \qquad (36)$$

where $T$ denotes the total number of trading periods and $r_t$ is the reward as defined in (18). The *ARR* metric assesses the profitability of the algorithm.

The second metric is the *Sharpe ratio*, which was defined by [11] as follow:

$$SR = \frac{E_t[\rho_t - \rho_{RF}]}{\sqrt{var(\rho_t - \rho_{RF})}} \qquad (37)$$

where $\rho_t$ is the rate of return defined in (17) and $\rho_{RF}$ represents the rate of return of risk-free asset. Since we select cash as the risk-free asset, $\rho_{RF}$ is equal to zero in the experiment. The *Sharpe ratio* mainly represents the risk-adjusted return of strategies.

---
[1] https://finance.yahoo.com/

In order to assess the risk resistance of an investment strategy completely, we introduce *Maximum Drawdown* [12] as the third metric. The formula of Maximum Drawdown (MDD) is

$$MDD = max_{\beta > t} \frac{y_t - y_\beta}{y_t} \quad (38)$$

This metric denotes the maximum portfolio value loss from a peak to bottom.

*C. Result and Analysis*

The performance of trading strategy is compared with several strategies as listed below:
- Robust Median Reversion (RMR) [14]
- the Uniform Buy and Hold (BAH), a portfolio management approach simply equally spreading the total fund into the preselected assets and holding them without making any purchases or selling until the end [7]
- Universal Portfolios (UP) [15]
- Exponential Gradient (EG) [16]
- Online Newton Step (ONS) [17]
- Aniticor (ANTICOR) [18]
- Passive Aggressive Mean Reversion (PAMR) [19]
- Online Moving Average Reversion (OLMAR) [20]
- Confidence Weighted Mean Reversion (CWMR) [21]
- Uniform Constant Rebalanced Portfolios (CRP) [15][22]

Since we set zero commission fee for DQN algorithm, all of the strategies mentioned above are tested without commission fee.

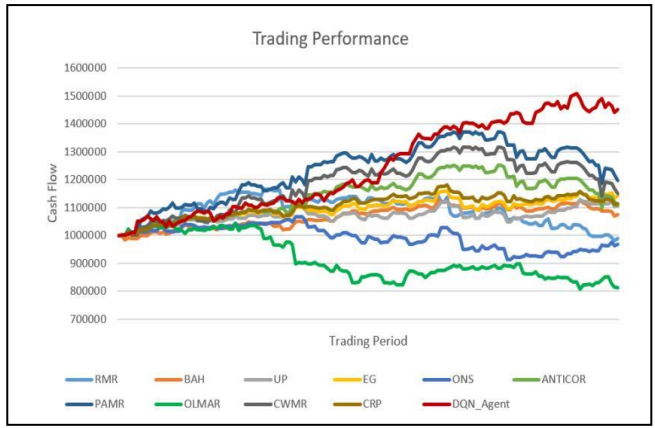

Figure 5. Trading Performance

Fig. 5 illustrates the accumulative return over the investment horizon of the test period as learning continuous from 2017/01/05 to 2017/11/17. Overall, the DQN strategy outperforms the benchmark strategies for the majority of the test trading period. Though the advantage is not quite obvious at the beginning, the DQN strategy tends to show strong benefits after the middle trading period, and the disparity between DNQ strategy and other benchmarks becomes obvious especially in the last few trading periods. Compared with OLMAR and ONS benchmarks which shows significant declines, the DQN strategy tends to steadily increase over the test trading period and the accumulative return is always above the initial cashflow which is 10000. Even though several benchmarks such as PAMR, SWMR and ANTICOR presents a moderate growth for a long time, these benchmarks are unable to continuously perform well in the last few trading periods when the DQN strategy shows a good performance.

Table I presents the comparison between DQN strategy and benchmarks in three aspects, namely Accumulative Rate of Return (ARR), Sharpe Ratio (SR) and Max Drawdown (MDD). It is obvious that the numerical results of DQN strategy are the best among other benchmarks in all aspects. In terms of ARR, the result of DQN strategy (45.05%) is more than twice of the second largest benchmarks PAMR (19.63%) over the test trading period. As for the risk measure, the DQN strategy still show the best performance by holding the minimum MDD (4.35%), comparing with the PAMR benchmark (13.46%) which is much higher. In aspect of Sharpe Ratio, DQN strategy (23.07%) is nearly twice of the second and third largest values which are from benchmark EG (12.63%) and CRP (11.00%). Even though EG and CRP performs nearly as good as DQN strategy in MDD, they show much lower value in the ARR with 14.20% and 11.30% respectively.

Overall, this result demonstrates the good profitability of DQN framework in comparison with traditional strategies.

TABLE I.  COMPARISON OF DIFFERENT ALGORITHMS

| | Rate of Return | SharpeRatio | Maximun Drawdown |
|---|---|---|---|
| RMR | -1.24% | -0.26% | 15.78% |
| BAH | 7.32% | 7.94% | 4.96% |
| UP | 11.38% | 9.51% | 6.48% |
| EG | 14.20% | 12.63% | 5.81% |
| ONS | -3.19% | -1.74% | 14.34% |
| ANTICOR | 10.44% | 6.87% | 12.21% |
| PAMR | 19.63% | 9.61% | 13.46% |
| OLMAR | -18.80% | -10.01% | 22.50% |
| CWMR | 14.87% | 7.81% | 13.45% |
| CRP | 11.30% | 11.00% | 5.92% |
| DQN_Agent | 45.05% | 23.07% | 4.35% |

X.  CONCLUSIONS AND FUTURE WORK

In this paper, we propose a deep reinforcement learning algorithm for portfolio management based on a discrete action space. We define a method to discretize the market action and combine this method with DQN algorithm. Five low-related US stocks are selected as experimental data and use accumulative rate of return, Sharpe ratio, Maximum Drawdown to compare the performance of our algorithm with 10 traditional strategies in the back testing set. The results show that this deep reinforcement learning algorithm is more profitable than all the surveyed traditional strategies, and it is also the least risky investment method on the back testing set we chose.

The limitation of our model is as follows. First, we set transaction cost as 0 (Section III, Assumption 4), so the profitability may be affected after the transaction cost is taken into consideration. Second, we assume that the volumes of stocks are large enough (Section III, Assumption 5) so each stock is available on any trading day. However, the stock might not be available sometimes, which will influence the profit as well.

For future work, we shall look into a DQN model with transaction cost. The trading in financial markets has a little transaction cost which may outweigh the profit in some transactions. In order to reduce the impact of transaction fees on agent's portfolio, we will try to increase the number of portfolio divisions (section IV), which means the least trading unit of the portfolio will shrink and transaction fees will be reduced to some extent. However, smaller trading unit leads to larger action space, which requires the agents should be able to explore and learn large-scale discrete action space effectively. Therefore, we will try to improving the exploration method of the agent such as Information-Directed Exploration [13]